\documentclass[aps,preprint,superscriptaddress,amsmath,amssymb,showpacs]{revtex4}

\usepackage{amsmath}
\usepackage{graphicx}
\usepackage{amssymb}
\usepackage{dcolumn}
\usepackage{bm}

\newcommand{\e}{\mathrm{e}}
\makeatletter

\begin{document}

\title{Biased diffusion in confined media: Test of the Fick-Jacobs approximation and
validity criteria}

\author{P. S. Burada}
\affiliation{Institut f\"ur Physik,
  Universit\"at Augsburg,
  Universit\"atsstr. 1,
  D-86135 Augsburg, Germany}

\author{G. Schmid}
\affiliation{Institut f\"ur Physik,
  Universit\"at Augsburg,
  Universit\"atsstr. 1,
  D-86135 Augsburg, Germany}

\author{D. Reguera}
\affiliation{Departament de F\'isica Fonamental,
  Facultat de F\'isica,
  Universidad de Barcelona,
  Diagonal 647, E-08028 Barcelona, Spain}

\author{J. M. Rub\'i}
\affiliation{Departament de F\'isica Fonamental,
  Facultat de F\'isica,
  Universidad de Barcelona,
  Diagonal 647, E-08028 Barcelona, Spain}

\author{P. H\"anggi}
\affiliation{Institut f\"ur Physik,
  Universit\"at Augsburg,
  Universit\"atsstr. 1,
  D-86135 Augsburg, Germany}

\date{\today}

\begin{abstract}

We study biased, diffusive transport of Brownian particles through
narrow, spatially periodic structures in which the motion is
constrained in lateral directions. The problem is analyzed under the
perspective of the Fick-Jacobs equation which accounts for the effect of
the lateral confinement by introducing an entropic barrier in a one
dimensional diffusion.
The validity of this
approximation, being based on the assumption of an instantaneous
equilibration of the particle distribution in the cross-section of
the structure,  is analyzed by comparing the different time scales
that characterize the problem. A validity criterion is  established
in terms of the shape of the structure and of the applied force. It
is analytically corroborated and verified by numerical simulations
that the critical value of the force up to which this description
holds true scales as the {\it square} of the periodicity of the
structure. The criterion can be visualized by means of a diagram
representing the regions where the Fick-Jacobs description becomes
inaccurate in terms of the
 scaled force versus the periodicity of the structure.

\end{abstract}


\maketitle

\section {Introduction }

In many transport phenomena such as those taking place in biological
cells, ion channels,  nano-porous materials and microfluidic devices
etched with grooves and chambers, Brownian particles, instead of
diffusing freely in the host liquid phase, undergo a constrained
motion. The uneven shape of these structures regulates the transport
of particles yielding important effects exhibiting peculiar
properties. The results have implications in processes such as
catalysis, osmosis and particle separation
\cite{hille,zeolites,Chou,liu,siwy,berzhkovski,kettner,muller,ai2006}
and  for the noise-induced transport in periodic potential
landscapes that lack reflection symmetry (ratchet systems) \cite{BM,
PT,RH} or ratchet transport mechanisms which are based on asymmetric
 geometries, termed "entropic" ratchet devices \cite{BM,
PT,RH,entropicR}. For example, it has been found that the separation
of DNA fragments in narrow channels \cite{Austin,Nixon,Chang} is
largely influenced by their shape. The translocation of structured
polynucleotides through nanopores also allows one to determine their
sequence and structure \cite{Gerland,Bundschuh,Keyser}.

The motion of the particles through these quasi-one-dimensional
structures can in principle be analyzed by means of the standard
protocol of solving the Smoluchowski equation with the appropriate
boundary conditions imposed. Whereas this method has been very
successful when the boundaries of the system possess a regular
shape, the challenge to solve the boundary value problem in the case
of uneven boundaries  represents typically a very difficult task. A
way to circumvent this difficulty consists in coarsening the
description by reducing the dimensionality of the system, keeping
only the main direction of transport, but taking into account the
irregular nature of these boundaries by means of an entropic
potential. The resulting kinetic equation for the probability
distribution, the Fick-Jacobs (F-J) equation, is similar in form to
the Smoluchowski equation, but now contains an entropic term.
The entropic nature of this term leads to a genuine dynamics
which is very different from that observed when the potential
has an energetic origin \cite{Reguera_PRL}.
It has been shown that the F-J equation can provide a very accurate
description of entropic transport in  2D and 3D channels of varying
cross-section \cite{Reguera_PRL, Reguera_PRE}.

However, the derivation of the F-J equation entails a tacit
approximation: The particle distribution in the transverse direction
is assumed to equilibrate much faster than in the main
(unconstrained) direction of transport. This equilibration
 justifies the coarsening of the description leading in turn to a simplification
of the dynamics, but raises the question about its validity when an
{\it external force is applied}. To establish the validity criterion
of a F-J description for such biased diffusion in confined media is,
due to the ubiquity of this situation, a subject of primary
importance.

Our objective with this work is to investigate in greater detail the
F-J approximation for biased diffusion and to set up a corresponding
criterion describing its regime of validity. We will analyze the
biased movement of Brownian particles in 2D and 3D periodic channels
of varying cross section and formulate different criteria for the
validity of such a F-J description.

The paper is organized in the following way: In
Sec.~\ref{sec:system}, we describe the physical situation and
introduce the model defined through the corresponding Langevin and
Fokker-Planck equations. In Sec.~\ref{sec:FJ}, we introduce the F-J
approach for the unbiased situation and extend it to the driven
case. Sec.~\ref{sec:validity} is devoted to establish a criteria for
the validity of the F-J approximation derived by comparing the
different characteristic time scales of the process. In
Sec.~\ref{sec:comparison}, the accuracy of the F-J description is
tested against numerical simulations for a 2D periodic channel, and
the conditions of validity are summarized in a diagram in terms of
the scaled force versus the periodicity of the structure. In
Sec.~\ref{sec:discussion}, we provide further explanations on when
and why the equilibration assumption fails and the F-J approach
leads to useable results. Finally, in Sec.~\ref{sec:conclusion} we
present our main conclusions.

\section{Diffusion in confined systems}
\label{sec:system}

\begin{figure}[t]
  \centering
  \includegraphics{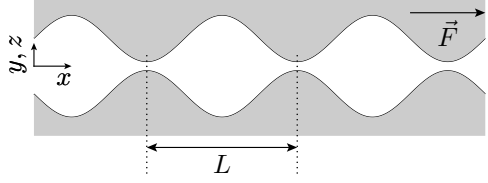}
  \caption{Schematic diagram of the channel confining the motion of the biased Brownian particles.
   The half-width $\omega$ is a periodic function of $x$ with periodicity $L$.} \label{fig:tube}
\end{figure}

In typical transport processes through pores or channels (like the
one depicted in Fig. \ref{fig:tube}), the motion of the suspended
particles is induced by application of an external potential $V(\vec
r)$ resulting in a force $\vec{F}$. In general, the dynamics of the
suspended Brownian particles is governed by Langevin's equation:

\begin{align}
\label{eq:newton}
m\ddot{\vec r}(t)= -\eta \dot{\vec r}(t)-\vec{\nabla}V(\vec r(t))+ \sqrt{\eta \, k_\mathrm{B}\,T}\,
\vec{\xi}(t)\, ,
\end{align}
where $\vec r$ is the two or three dimensional position vector of a particle of mass $m$,
$\eta$ is its friction coefficient, $k_{\mathrm{B}}$ the Boltzmann constant, $T$ the temperature,
 and a dot over the quantity refers to time derivative.
The thermal fluctuations due to the coupling of the particle with
the environment are modeled by a zero-mean Gaussian white noise
$\vec{\xi}(t)$, obeying the fluctuation-dissipation relation
$\langle \xi_{i}(t)\,\xi_{j}(t') \rangle = 2\, \delta_{ij}\,
\delta(t - t')$ for $i,j = x,y,z$. In the over-damped case, i.e.
when $m\ddot{\vec r}(t)<<\eta\dot{\vec r}(t)$, the inertia term in
Eq.~\eqref{eq:newton} can safely neglected and the Langevin equation
describing the dynamics of a Brownian particle within the channel
reads:

\begin{align}
  \label{eq:langevin}
  \eta\, \frac{\mathrm{d}\vec{r}}{\mathrm{d} t} = -\vec{\nabla}V(\vec r(t)) +
  \sqrt{\eta \, k_\mathrm{B}\,T}\, \vec{\xi}(t)\, .
\end{align}

In addition to Eq.~\eqref{eq:langevin}, the full problem is set up by imposing reflecting
boundary conditions at the channel walls.

The corresponding Fokker-Plank equation for the time evolution of
the probability distribution $P(\vec r, t)$ takes the form
\cite{Risken,hanggithomas}:
\begin{align}
\label{eq:fp}
\frac{\partial P(\vec r, t)}{\partial t} = - \vec{\nabla} \vec{J}(\vec{r},t)\, ,
\end{align}
where $\vec J(\vec r, t)$ is the probability current:
\begin{align}
  \label{eq:particlecurrent}
  \vec J(\vec r, t) =
  - \left( \frac{\vec{\nabla} V(\vec r)}{\eta} +
    D_0 \vec{\nabla}\right) P(\vec r, t) \, ,
\end{align}
and

\begin{align}
  D_0=k_{\mathrm{B}}T/\eta \;,
\end{align}
\noindent
denotes the diffusion coefficient of the suspended particles.

Due to the impenetrability of the channel walls, the normal component of the probability current $\vec J(\vec r, t)$
 vanishes at the boundaries.
If $\vec{n}$ denotes the vector perpendicular to the channel walls, the reflecting boundary conditions read:
\begin{align}
  \label{eq:bc-general}
  \vec J(\vec r, t) \cdot \vec{n}=0\, .
\end{align}

In this paper we focus on the case of a symmetric 2D channel where the force is constant and directed
along the axis, cf. Fig.~\ref{fig:tube}. The half-width of the 2D channel is given by a periodic function $\omega(x)$,
i.e. $\omega(x+L)=\omega(x)$ for all $x$. In this case, the boundary condition reads
\begin{align}
  \label{eq:bc}
  - \frac{\mathrm{d} \omega (x)}{\mathrm{d} x}
 \left[ \frac{F}{\eta} P(x,y,t) - D_0\frac{\partial P(x,y,t)}{\partial
     x} \right]
  - D_0\frac{\partial P(x,y,t)}{\partial y} = 0\, ,
  \end{align}
at $y=\pm \omega (x)$. For an arbitrary form of $\omega (x)$, the
boundary value problem defined through Eqs.~\eqref{eq:fp},
\eqref{eq:particlecurrent}  and \eqref{eq:bc} is very difficult to
solve. Despite the inherent complexity of this problem an
approximate solution can be found  by introducing an effective
one-dimensional description where geometric constraints and
bottlenecks are considered as {\itshape entropic barriers}
\cite{Reguera_PRL,Jacobs,Zwanzig,Reguera_PRE,Percus,ai2006}.

\section{The Fick-Jacobs approximation}
\label{sec:FJ}

In the absence of an external force, i.e. when $\vec{F}=0$, it was shown \cite{Jacobs,Zwanzig} that
the dynamics of particles in confined structures (such as that of Fig.~\ref{fig:tube})
can be described by the F-J equation
\begin{align}
\label{eq:fickjacobs}
\frac{\partial P(x,t)}{\partial t}=\frac{\partial}{\partial
  x}\left(D_{0}h(x)\frac{\partial}{\partial
    x} \frac{P}{h(x)}\right) \, ,
\end{align}
obtained from the 3D (or 2D) Smoluchowski equation after elimination
of $y$ and $z$ coordinates. Here $P(x,t)$ is the probability
distribution function along the axis of the 2D or 3D channel,
$h(x)$ is the dimensionless transverse cross-section $h(x):= \pi\,
[\omega(x)/L]^{2} $ in 3D, and the dimensionless width $h(x):=
2\,\omega(x)/ L$ in 2D, where $\omega(x)$ is the radius of the channel
in 3D (or the half-width of the channel in 2D). This description is
 valid for $\left|\omega'(x)\right|\ll1$ (the prime
refers to the first derivative, i.e.
$\omega^{\prime}(x)=\mathrm{d}\omega(x) / \mathrm{d}x$), and it has been
shown that the introduction of an effective $x$-dependent diffusion
coefficient can considerably improve the accuracy of the kinetic
equation, thus extending its validity to more winding structures
\cite{Zwanzig,Reguera_PRE,Percus}. The expression:
\begin{align}
  \label{eq:diffusionconst}
  D(x)=\frac{D_{0}}{(1+\omega'(x)^{2})^{\alpha}}\, ,
\end{align}
where $\alpha=1/3,1/2$ for two and three dimensions, respectively,
has been shown to accurately account for the curvature effects
\cite{Reguera_PRE}.

In the presence of a constant force $F$ along the direction of the
channel the F-J equation can be recast into the expression \cite{Reguera_PRE,Reguera_PRL}

\begin{align}
\label{eq:fickjacobs_ours}
\frac{\partial P}{\partial t}=\frac{\partial}{\partial
  x}\left(D(x)\frac{\partial P}{\partial
    x}+\frac{D(x)}{k_\mathrm{B}\,T}\frac{\partial A(x)}{\partial x}P\right)\,
\end{align}
which defines the free energy
  $A(x) := E - T\, S = -F\, x - T\, k_\mathrm{B}\, \ln  h(x)$,
where $E = V = -Fx$ denotes the energy contribution and
$S=k_\mathrm{B} \,\ln h(x)$ the entropy contribution. For a
symmetric channel with periodicity $L$, the free energy assumes the
form of a periodic tilted potential.

Note that Eq.~\eqref{eq:fickjacobs_ours} may typically describe the
time-evolution of a  1D particle distribution within an energy
landscape. In the present context, however, due to the reduction of
the geometric confinements in 2D or 3D space into one dimension, we
end up with an entropic contribution to the free energy. In absence
of a fixed bias $F = 0$, we deal with a pure entropic situation with
the free energy reading $A(x)=-T\,k_{\mathrm{B}}\, \ln h(x)$.
Equation~\eqref{eq:fickjacobs_ours} then corresponds to a F-J
equation with a spatial-dependent diffusion coefficient. Likewise,
for a situation involving solely transport in a one-dimensional
energy landscape, the free energy reduces to  $A(x)=V(x)$ with
$d\omega (x)/dx =0$, yielding to the well-known Fokker-Planck
equation in 1D, with a constant diffusion coefficient $D_0$, reading
\begin{align}
  \label{eq:fp1D}
  \frac{\partial P}{\partial t}=\frac{\partial}{\partial x}\left(\frac{V^{\prime}}{\eta}P + D_0
\frac{ \partial P}{ \partial x}\right)\, .
\end{align}

Recently, we have shown that the dynamics of a confined Brownian
particle, in presence of an applied bias, can accurately be
described by means of equation \eqref{eq:fickjacobs_ours}
\cite{Reguera_PRL}. In the presence of very strong applied bias, and
for more winding structures, however, the F-J equation becomes
inaccurate. In the present work we present further numerical and
analytical results and will set up tailored criteria under which the
F-J approximation assumes good validity.

\subsection{Particle current}

One of the key quantities in transport through quasi-one-dimensional structures is the study of the average
particle current $\langle \dot{x} \rangle$,
or equivalently the nonlinear mobility, which is defined as the ratio of the average particle current
and the applied force $F$.
For the average particle current we derive an expression which is
similar to the Stratonovich formula for the current in titled periodic
energy landscapes \cite{stratonovich,HTB,Reimann,Heinsalu}, but
with a spatial diffusion coefficient.
A detailed derivation of this expression is given in the
Appendix~\ref{appendix}, cf. Eq.~\eqref{eq:j9}.
Hence, we obtain the nonlinear mobility for a 2D channel with a shape defined by $\omega(x)$:
\begin{align}
  \label{eq:mobility1}
   \mu(F,L,\beta)\, \eta :=  \frac{\langle \dot{x} \rangle }{F}\, \eta
   = \frac{1}{ \beta F} 
  \frac{ 1-\exp(-\beta FL)}{ \int_{0}^{L}\frac{\mathrm{d}x}{L} \e^{-\beta A(x)} \int_{x}^{x+L}\mathrm{d}x^\prime
    \left[ 1 +
      \left( \frac{\mathrm{d} \omega(x^{\prime})}{ \mathrm{d}x^{\prime}}\right)^2 \right]^{\alpha}
    \e^{\beta A(x^\prime)}
} \, ,
\end{align}
where we have made use of Eq.~\eqref{eq:diffusionconst} and the relations $D_0=k_{\mathrm{B}}T/\eta
$ and $\beta=1/k_{\mathrm{B}}T$. Substituting $z^{\prime}=x^{\prime}/L$ and $z=x/L$, the
nonlinear
mobility scaled with the
friction coefficient $\eta$ can be expressed, for the case of
constant forcing, in terms of a single, dimensionless scaling
parameter \cite{Reguera_PRL}:
\begin{align}
  \label{eq:scalingparameter}
  f:=\beta FL\, .
\end{align}
Therefore, Eq.~\eqref{eq:mobility1} for $A(x)=-Fx-\beta^{-1}\ln(h(x))$ leads to
\begin{subequations}
  \label{eq:mobility}
  \begin{align}
    \label{eq:mobility2}
    \mu(f)\, \eta = \frac{1}{f} \, \frac{1-\exp(-f)}{\int_0^1 \mathrm{d}z I(z,f)}\,  ,
    \end{align}
    where
    \begin{align}
      \label{eq:mobility3}
      I(z,f) = \exp(fz) h(Lz) 
      \int_z^{z+1} \mathrm{d} z^{\prime} \exp(-f z^{\prime}) h^{-1}(L z^{\prime})
      \left[ 1 + \left(\frac{1}{L}\frac{\mathrm{d}\omega(Lz^{\prime})}{\mathrm{d} z^{\prime}}\right)^{2}\right]^{\alpha} \, .
    \end{align}
\end{subequations}
Eq.~\eqref{eq:mobility} can be transformed into Eq.~(6) in Ref.~\cite{Reguera_PRL} by interchanging the
order of integration.

The asymptotic values of the nonlinear mobility for the cases $f\to
0$ and $f\to\infty$, can be evaluated to read:
\begin{align}
  \label{eq:limf0}
  \lim_{f\to0} \mu(f)\, \eta = \frac{1}{\left<h(x)\right>\left<\frac{D_{0}}{D(x)}h^{-1}(x)\right>}\,  ,
\end{align}
and

\begin{align}
  \label{eq:limfinf}
 \lim_{f\to\infty}  \mu(f)\, \eta = \frac{1}{\left<\frac{D_{0}}{D(x)}\right>}\,  ,
\end{align}
where
\begin{align}
\langle
g(x)\rangle=\frac{1}{L}\int_{0}^{L}g(x)\mathrm{d}x\,,\label{eq:meancurvesquared}
\end{align}
is the average over a spatial period, given an arbitrary periodic function $g(x)$.

\section{Validity of the Fick-Jacobs description in presence of a constant bias}
\label{sec:validity}

The reduction of dimensionality done implicitly in the formulation of the F-J equation relies
on the assumption of equilibration in the transverse direction.
An estimate of the conditions under which equilibration occurs
can be made by analyzing the
different time scales involved in the problem. For the sake of simplicity,
let us focus on the situation of a 2D channel, although the same discussion
can readily be extended to 3D. In a 2D channel in the presence of an
external force, in the axial direction, one can identify three characteristic processes,
with the corresponding different time scales. One is diffusion
in the transverse direction over a distance $\Delta y$, whose time
scale is \begin{align}
\tau_{y}=\frac{\Delta y^{2}}{2D_{0}}\,.\label{eq:taudiffy}\end{align}

Similarly, the time scale associated to diffusion in the axial direction
is

\begin{align}
\tau_{x}=\frac{\Delta x^{2}}{2D_{0}}\,.\label{eq:taudiffx}
\end{align}

The third time scale is defined through the characteristic time associated to the drift
(ballistic motion) over a distance $\Delta x$ given by \begin{align}
\frac{\Delta x}{\tau_{drift}}\sim\frac{F}{\eta}=\frac{fD_{0}}{L}\,,\label{eq:taudriftx_pr}\end{align}
 where we have used the scaling factor $f=FL/k_{\mathrm{B}}T$. Rearranging
the previous expression, we obtain \begin{align}
\tau_{drift}=\frac{L\Delta x}{D_{0}f}\,.\label{eq:taudriftx}\end{align}

In order to have a good equilibration in the transverse direction,
the characteristic time scale associated to diffusion in this direction
has to be much smaller than the other two time scales. Therefore,
in the absence of an external force, equilibration in the
transverse direction occurs if $\tau_{y}/\tau_{x}\ll1$. This results
in the condition: \begin{align}
\frac{\Delta y^{2}}{\Delta x^{2}}\sim\omega^{\prime}(x)^{2}\ll1\,,
\label{eq:FJcriteria}
\end{align}
which constitutes the
validity criterion of the F-J approach such as put forward by Zwanzig
\cite{Zwanzig}.

In presence of a force along the axis, equilibration
in the transverse direction demands that the condition $\frac{\tau_{y}}{\tau_{drift}}\ll1$
also holds. Consequently,
\begin{align}
\frac{f\Delta y^{2}}{2L\Delta
  x}\sim\frac{2f\omega(x)^{2}}{L^{2}}\ll1\,,\label{eq:driftcriteria}
\end{align}
 where in the second step we have replaced the characteristic distances
$\Delta y$ by the width $2\omega(x)$, and $\Delta x$ by $L$.

A general estimate of the criteria that has to be satisfied is that
$\mathrm{max}(\frac{\tau_{y}}{\tau_{x}},\frac{\tau_{y}}{\tau_{\mathrm{drift}}})\ll1$.
An even stronger criteria in order for the F-J description to
hold in presence of a constant force can be put forward by considering
the sum of the two ratios, cf. Eqs.~\eqref{eq:FJcriteria} and \eqref{eq:driftcriteria},
i.e.:
\begin{align}
\omega^{\prime}(x)^{2}+\frac{2f\omega(x)^{2}}{L^2}\ll1\label{eq:criterialocal}
\end{align}

Eq.~\eqref{eq:criterialocal} provides a quite stringent criteria
that indicates when the F-J description of a system is expected
to be valid. Note also that this is a \textit{local} criterion, i.e.
for a given channel, there would be regions (associated to drastic
changes in the shape of the channel, i.e $\omega^{\prime}(x)^{2}\gg1$
) where  equilibration in the transverse direction is not feasible,
whereas in others is fulfilled. It is then more
convenient to work with a \textit{global criterion} of validity rather
than with a local one. One way of getting that global condition is by averaging
the local criterion over the period $L$ of the channel, yielding
one of our main results:
\begin{align}
\langle
\omega^{\prime}(x)^{2}\rangle+\frac{2f}{L^2}\langle\omega(x)^{2}\rangle\ll1\,.
\label{eq:criteriaglobal}\end{align}

In order to get an explicit estimate of the dependence of the maximum force
value on the periodicity $L$ of the channel, we define a critical
force value $f_{c}$, for which the inequality~\eqref{eq:criteriaglobal}
becomes an equality, i.e. $\langle
\omega^{\prime}(x)^{2}\rangle+\frac{2f_{c}}{L^2}\langle\omega(x)^{2}\rangle=1$.
Then the critical force value reads:
\begin{align}
f_{c}=\frac{L^{2}}{2\langle\omega(x)^{2}\rangle}\left(1-\langle\omega^{\prime}(x)^{2}\rangle\right)
\, ,\label{eq:criticalforcevalue}
\end{align}
 which indicates that the critical force scales asymptotically as $L^2$, if we fix the overall shape of
 the channel and change only its periodicity.

Eq.~\eqref{eq:criticalforcevalue} provides an estimate of the minimum
forcing beyond which the F-J description is expected to fail
in providing an accurate description of the dynamics of system. The quantitative value of the critical
force will obviously depend on the level of accuracy sought. What  is really important is how it depends
(or scales) with the relevant parameters of the problem.

\section{Numerical simulations for a 2D channel}
\label{sec:comparison}

In order to check the consistency of the criteria proposed and the validity of the F-J
description in the presence of a force, we will compare the analytical result for the
scaled nonlinear mobility obtained in  Eq.
 \eqref{eq:mobility} with the simulation results of the overdamped
 Langevin dynamics in
 Eq.~\eqref{eq:langevin}
for the 2D periodic channel sketched in Fig.~\ref{fig:tube}. We
remark here that  the extension of our scheme to 3D with a
rotational symmetry along the transport axis is possible as well.
This will consume more computation time, but the overall findings
remain qualitatively robust. This feature has been verified with a
few numerical tests.

The shape of the channel is described by

\begin{align}
\omega(x)=a\sin(2\pi x/L)+b\,,\label{eq:boundary}
\end{align}
where $a$ controls the slope of the channel walls and $2(b-a)$ gives
the channel width at the bottlenecks.
In order to use dimensionless
quantities we refer any physical length to $a$, i.e. the
scaled boundary condition then reads $\omega(x) = \sin(2\pi x/L) + b$,
where $b$ is now a dimensionless quantity. In our case, we have used $b=1.02$.
\begin{figure}[t]
\centering
  \includegraphics{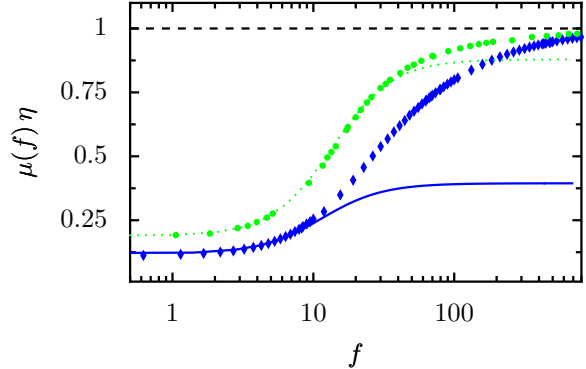}
  \caption{(color online) The numerically simulated (symbols) and
    analytically calculated, cf. Eq.~\eqref{eq:mobility}, (lines)  dependence of the scaled nonlinear
    mobility $\mu(f)\eta$ {\it vs.} the scaling parameter $f=FL/k_\mathrm{B}T$ is
    depicted for two channels in 2D with different spatial periods.
    For both channels the scaled half width is given by $\omega(x)=\sin(2\pi x/L) + 1.02$;
    $L=1:$ diamonds and solid line(blue), $L=2\pi:$ circles and dotted line (green).
    The dashed line indicates the deterministic limit $\mu(f)\eta =
    {\langle\dot x \rangle}/(F / \eta) = 1$.}
\label{fig:mobility}
\end{figure}
The behavior of the quantities of interest have been corroborated
by Brownian dynamic simulations performed by  integration
of the Langevin equation using the standard
stochastic Euler-algorithm. The average particle current in $x$-direction has been derived
from an ensemble-average  of about $3\cdot 10^{4}$ trajectories according to the following expression:

\begin{align}
\label{eq:current-num}
\langle \dot{x}\rangle=\lim_{t\to \infty} \langle \frac{x(t)}{t} \rangle \, .
\end{align}

In order to test the accuracy of the F-J description, we have
evaluated the behavior of the nonlinear mobility as a function of
the scaled force $f$, for different values of the periodicity $L$.
In Fig.~\ref{fig:mobility}, the dependence of the scaled nonlinear
mobility on  $f$ is depicted for two different  periodic structures
of spatial period $L$, and for $b=1.02$. In both cases, the
nonlinear mobility depicts a monotonic increase with increasing the
force. The value of the scaling parameter $f =\beta F L$ up to which
the F-J approximation with spatial dependent diffusion coefficient
$D(x)$ provides an accurate description depends on the spatial
period $L$. Consistent with the F-J approximation scheme, for rather
smoothly varying cross-section, or equivalently large periods $L$,
the agreement between our precise numerics and the analytic solution
is attained for quite larger $f$-values as compared to the case with
strongly winding structures or small periods. The analytic solution
does not capture the correct limiting value for large $f$. Whereas
the analytic result of the F-J equation tends to the value given by
Eq.~\eqref{eq:limfinf}, the limiting value observed in the
simulations always tends to $1$  which corroborates with the
deterministic limit. That indicates that in the case of very strong
forcing, the particles almost travel ballistically, and thus do not
feel the effects of the boundaries. This all is consistent with the
accompanying breakdown of the F-J approximation in this strong
forcing limit for which the deterministic dynamics dominates the
transport.

\begin{figure}[t]
  \centering
  \includegraphics{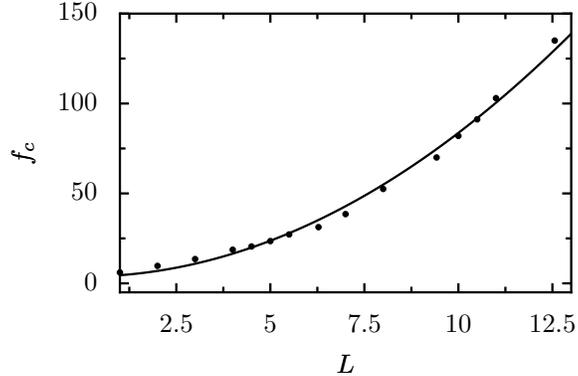}
  \caption{The dependence of the critical value of the scaling
parameter, i.e. $f_c$, on the periodicity $L$ of a 2D channel
defined by the dimensionless boundary function $\omega(x)=\sin(2\pi
x/L)+1.02$ is depicted. For $f<f_c$ the relative error of the
Fick-Jacobs description is less than $1\%$. The solid line is a
$L^2$-fit of the critical values obtained by comparison of the
approximative analytic and the exact numerical results.}
  \label{fig:criteria}
\end{figure}

By comparing the exact numerical results with the analytic solution of the F-J equation,
Eq. \eqref{eq:mobility}, we can identify a critical value of the scaled force, $f_c$, beyond which
the relative error in the mobility exceeds a certain value. This critical force plays a similar role to
that introduced  in the previous section.

For our example of a 2D channel whose
shape is defined by Eq. \eqref{eq:boundary}, the validity criterion given
by Eq. \eqref{eq:criticalforcevalue} simplifies to

\begin{align}
f_{c}=\frac{L^2}{1+2b^2}\left(1-\frac{2\pi^{2}}{L^2}\right)\,.
\label{eq:criticalforcevalueourexample}
\end{align}
thus predicting that the  critical value of the force scales as $L^{2}$.

This prediction has been verified by the simulations. Fig. \ref{fig:criteria} shows the value of the
critical force for a tolerance of $1\%$ as a function of the periodicity $L$.
The critical value of the force depends {\it quadratically} on the
periodicity $L^2$, as predicted.

\begin{figure}[t]
  \centering
  \includegraphics{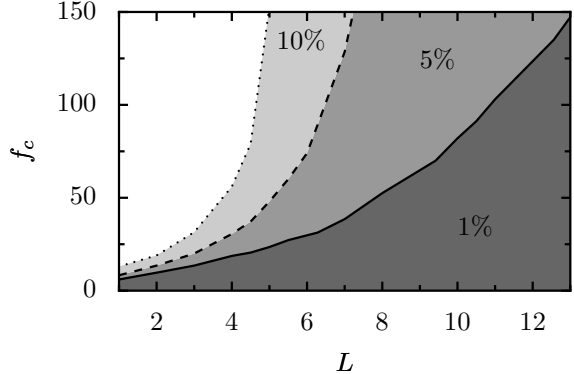}
  \caption{ The {\itshape validity diagram} of the Fick-Jacobs approximation for biased diffusion
   is obtained upon a comparison between the
    precise numerics with the approximative analytic solution,
    cf. Eq.~\eqref{eq:mobility} for a 2D channel with boundary function $\omega(x)=\sin(2\pi
    x/L)+1.02$. The dependence of the critical value
    of the scaling parameter on the periodicity is depicted for three
    different relative errors; $1\%$: solid line, $5\%$: dashed line
    and $10\%$: dotted line. Below these limiting lines the analytic treatment agrees
    within the corresponding prescribed  relative error. }
 \label{fig:phase}
 \end{figure}

In Fig. \ref{fig:phase} we illustrate, for the considered
two-dimensional channel, the regions where the F-J approximation is
accurate when compared with the  simulations. We depict the
dependence on the periodicity $L$ of the maximal critical scaled
force, obtained by comparing numerical results with the analytic
solution for the particle current, for different required relative
errors. This diagram shows the regions in parameter space in terms
of $f$ and $L$ for which an accurate solution is provided. Thus, it
is possible to provide an accurate result by using the analytic
solution over a wide range of the scaled parameter and the
periodicity.

\section {The hypothesis of equilibration in the transverse direction}
\label{sec:discussion}

\begin{figure}[t]
\centering
\includegraphics[width= 6.5cm ,height = 5cm, clip=true]{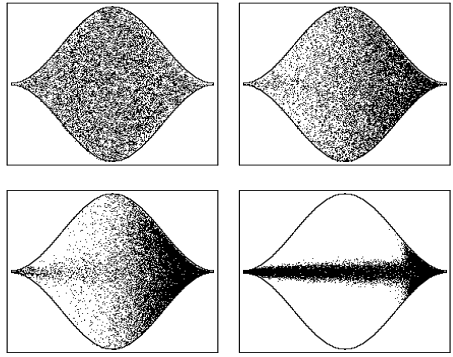}
 \caption{Steady-state particle distribution mapped into a single period of the 2D channel defined
by the boundary function $\omega(x)=\sin(2\pi x/L) + 1.02$ with $L=1$ for four different values of the
dimensionless scaling parameter $f$: upper left panel: $f=0.2$;
upper right panel: $f=3.0$; lower left panel: $f=7.0$;
lower right panel: $f=50$.}
\label{fig:snap}
\end{figure}

From our simulations, we can actually analyze the validity of the hypotheses of
equilibration in the transverse direction on which the F-J description relies.
In Fig.~\ref{fig:snap}, we show the steady state distribution of a certain
number of Brownian particles for different values of the scaled force $f$. As the force increases, we can
clearly see how the particles are not homogeneously distributed in $y$-direction, evidencing the failure of
the equilibration assumption. This effect is specially
dramatic in Fig. \ref{fig:snap}(lower right panel), where the force is so strong that
the particles do not fully explore the available space  in $y$-direction.

A more detailed analysis could be provided
by checking the normalized steady-state probability distribution in the transverse direction
at a given  $x$-positions, i.e.
\begin{align}
  \label{eq:yprob}
  P_{x}^{\mathrm{st}}(y):=\frac{P^{\mathrm{st}}(x,y)}{\int_{-\omega(x)}^{\omega(x)} \, \mathrm{d}y\, P^{\mathrm{st}}(x,y) }\, .
\end{align}

\begin{figure*}[t]
\centering
  \includegraphics{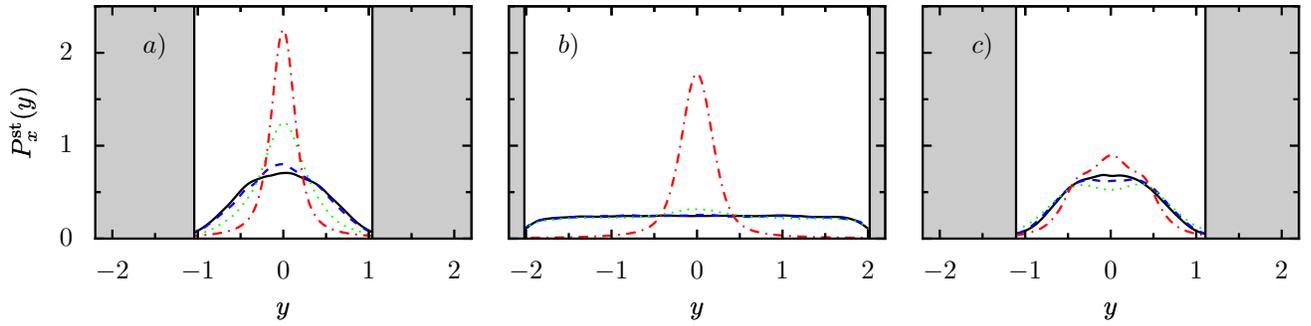}
  \caption{(color online) Normalized steady-state probability
 distribution of particles in $y$-direction $P_{x}^{\mathrm{st}}(y)$,
 cf. Eq.~\eqref{eq:yprob} for different values of the scaling parameter
 $f$ -- solid line (black): $f$ = 0.2; dashed line (blue): $f$ = 3.0;
 dotted line (green): $f$ = 7.0; dashdotted line (red): $f$ = 50.0, at various
 positions along the length of the channel with the boundary function
 $\omega(x)=\sin(2\pi x/L) + 1.02$, $L=1$:
 $x/L=0.2$ (panel (a)), $x/L=0.5$ (panel (b)) and $x/L=0.8$ (panel
 (c)). The grey regions indicate the outside of the channel.}
  \label{fig:disty}
\end{figure*}

In Fig.~\ref{fig:disty}, we represent the
steady-state probability density at three different locations
along the channel, corresponding to $x/L = 0.2$, $x/L = 0.5$, and $x/L = 0.8$.
For small values of $f$, at
$x/L = 0.5$, where $\omega^{\prime}(x) = 0$, the $P_{x}^{\mathrm{st}}(y)$ is flat, indicating a perfect equilibration in the
transverse direction.
However, at $x/L = 0.2$, and $x/L = 0.8$, the system is not equilibrated and $P_{x}^{\mathrm{st}}(y)$
is bell-shaped. Notice that for $x/L = 0.2$ and $x/L = 0.8$, the
distributions are very much similar for small values of $f$, corroborating
that the equilibration depends on  $\omega^{\prime}(x)^2$, which is the same in both cases.

At large force strengths particles concentrate along the axis of the channel. In this situation,
the assumption of equilibration along the
transverse direction fails.
Particles would only feel the presence of the boundaries when they
are close to the bottlenecks.
Hence, in the limit of very large force values, the influence of the
entropic barriers practically disappears.
In this limit, the correction in the
diffusion coefficient
leading to a spatial dependency, i.e. $D(x)$, overestimates the influence of the
entropic barriers.

\begin{figure}[t]
\centering
\includegraphics{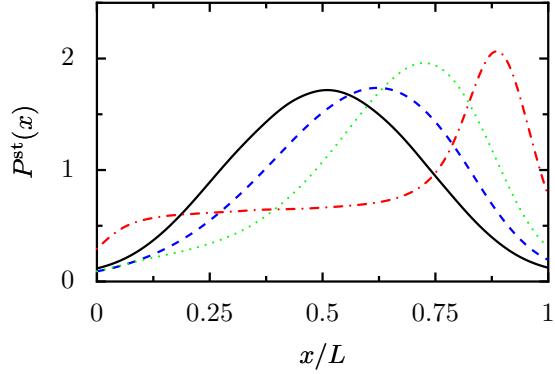}
 \caption{(color online) Normalized steady-state probability distribution of particles along the
length of the channel $P^{\mathrm{st}}(x)$, cf. Eq.~\eqref{eq:xprob},
for the corresponding set of scaling parameters $f$ as those chosen
in Fig.~\ref{fig:disty}.} \label{fig:distx}
\end{figure}

Fig.~\ref{fig:distx} depicts the normalized stationary distribution
function of particles in $x$-direction at various force strengths,
obtained from numerical simulations:
\begin{align}
  \label{eq:xprob}
  P^{\mathrm{st}}(x):=\frac{\int_{-\omega(x)}^{\omega(x)} 
    \,\mathrm{d}y
    P^{\mathrm{st}}(x,y)}{\int_{0}^{L}\mathrm{d}x\int_{-\omega(x)}^{\omega(x)} \,
    \, \mathrm{d}y\, P^{\mathrm{st}}(x,y) }\, .
\end{align}

At very low force strengths particles are evenly distributed, and the
probability distribution $P^{\mathrm{st}}(x)$ scales with the
cross-section of the channel.
When increasing the force, the maximum of $P^{\mathrm{st}}(x)$ is shifted
towards the exit of the cell and the particles accumulate {\itshape in
front of} the bottleneck.
This behavior can also be observed in
Fig.~\ref{fig:snap}.
In the large force regime
$P^{\mathrm{st}}(x)$ is almost constant over a wide range of
$x$-values, indicating a minor influence of the shape of the structure on the dynamics of the particles.
 In the limit of large $f$-values the
plateau extends and covers the whole period.
In this situation, a deterministic treatment of the problem
leads to adequate results.

\section {Conclusions}
\label{sec:conclusion}

With this work we have investigated the validity conditions under
which the Fick-Jacobs approximation provides an accurate description of the
biased diffusion of Brownian particles in 2D and 3D confined systems.
We have established a validity criterion formulated in terms of the
sinuosity of the channel $\omega(x)$, as done in the unbiased case
\cite{Jacobs,Zwanzig,Reguera_PRE,Percus}, and on the scaling parameter that
causes forced diffusion. This scaling parameter compares the work
done on a particle travelling  a distance equal to the spatial period $L$ of
the channel  with the available thermal energy.  Interestingly, the
critical value of this scaling parameter up to which Fick-Jacobs
equation holds depends on the {\it square} of the period. This
dependence follows from the analysis of the different time scales
that rule  the biased, diffusive dynamics. We have constructed a
validity diagram showing the region of parameters (spanned by $f$ and
$L$ ) in which  the Fick-Jacobs approximation describes the overdamped diffusive transport
accurately cf. Fig.~\ref{fig:phase}. We have also investigated  numerically
the conditions for a fast equilibration  in the transverse direction which is
vital for the accurateness of the
Fick-Jacobs approximation.  The results  presented evidence
the usefulness  of the Fick-Jacobs description with a spatially-dependent
diffusion coefficient at small applied bias. The obtainment of a simplified
Fick-Jacobs like description that covers also  for  intermediate-to-strong values of
the scaled force still presents an open challenge. The availability
of such a description would  be beneficial for the detailed
understanding of diffusive transport processes occurring in confined
media, the latter dictating the nonequilibrium transport behavior
 in a  great variety of systems far from thermal
equilibrium \cite{hille, zeolites, Chou, liu, siwy, berzhkovski,
kettner, muller, ai2006, BM, PT, RH, entropicR, Austin, Nixon,
Chang, Gerland, Bundschuh, Keyser, Reimann, Heinsalu}.

\section*{ACKNOWLEDGEMENTS}
This work has been supported by the
DGiCYT under Grant No BFM2002-01267 (D.R.),
ESF STOCHDYN project (G.S., D.R., J.M.R., P.H.), the Alexander von
Humboldt Foundation (J.M.R.),
the Volkswagen Foundation (project I/80424, P.H.),
the DFG via research center, SFB-486, project A10 (G.S., P.H.)
and via the project no. 1517/26-1 (P.S.B., P.H.), and by the Nanosystems Initiative Munich (P.H.).

\begin{appendix}
\section{Particle current}
\label{appendix}

The evolution equation of the probability distribution of overdamped Brownian particles
in 1D can also be expressed as

\begin{align}
  \label{eq:transportform}
  \frac{\partial P(x,t)}{\partial t}=\frac{\partial}{\partial
    x}\left(D(x)\, \e^{-\beta A(x)}\frac{\partial}{\partial
      x} \e^{\beta A(x)}\, \right) P(x,t)
\end{align}
where $D(x)$ is a position dependent diffusion coefficient,  and $A(x)$
is the free energy landscape for the reaction coordinate $x$.

Eq.~\eqref{eq:transportform} results from the
continuity equation:

\begin{align}
\label{eq:continuity}
\frac{\partial}{\partial t}P(x,t) = - \frac{\partial}{\partial x}
J(x,t) \, ,
\end{align}
with the probability current:
\begin{align}
\label{eq:j1}
 J(x,t)=\,-\, D(x)\, \e^{-\beta A(x)}\frac{\partial}{\partial x}
 \e^{\beta A(x)}\, P(x,t)\, .
\end{align}
In the case of a periodic diffusion coefficient,  $D(x+L)=D(x)$, and
a tilted periodic free energy, $A(x+L)=A(x)-FL$, it is convenient to
define the reduced probability density and the corresponding current,

\begin{align}
  \label{eq:redprobdens}
  \hat{P}(x, t) &= \sum_n P(nL+x, t) \, ,\\
  \hat{J}(x, t) &= \sum_n J(nL+x, t) \, ,~n \in \mathbb{Z}  \, .
\end{align}

By definition these functions are periodic with  periodicity $L$,
$\hat{P}(x+L, t) = \hat{P}(x, t)$ and $\hat{J}(x+L, t) = \hat{J}(x,
t)$. Moreover $\hat{P}(x, t)$ and $\hat{J}(x,
t)$ enter a continuity equation~\eqref{eq:continuity} and
$\hat{P}(x,t)$ is normalized on any interval $(x, x + L)$, provided that
$P(x, t)$ is normalized, e.g. $\int_{-\infty}^{+\infty} \mathrm{d}x \, P(x, t) = 1$.
In the {\it steady state} limit the probability current is a constant,
$\hat{J}(x,t) \rightarrow \hat{J}$, thus Eq.~\eqref{eq:j1} becomes

\begin{align}
  \label{eq:j2}
 \hat{J}=\,-\,D(x)\, \e^{-\beta A(x)}\frac{\partial}{\partial x}
 \e^{\beta A(x)}\, \hat{P}^{\mathrm{st}}(x)\, .
\end{align}

Multiplying both sides of Eq.~\eqref{eq:j2} by $1/D(x) \e^{ \beta
  A(x)}$, and integrating over a period  we obtain
\begin{align}
  \label{eq:j3}
  \hat{J}\int_{x}^{x+L} \frac{1}{D(x^\prime)}\, \e^{\beta A(x^\prime)}\,
  \mathrm{d}x^\prime  = 
  \int_{x}^{x+L} \frac{\partial}{\partial x^\prime} \e^{\beta
    A(x^\prime)}\, \hat{P}^{\mathrm{st}}(x^\prime) \mathrm{d} x^\prime \, .
\end{align}
Using now the
conditions $\hat{P}^{\mathrm{st}}(x+L) = \hat{P}^{\mathrm{st}}(x)$ and
$A(x+L) = A(x) - \beta F L$,  Eq.~\eqref{eq:j3} reduces to
\begin{align}
\label{eq:j5}
\hat{J}\int_{x}^{x+L} \frac{1}{D(x^\prime)}\, \e^{\beta A(x^\prime)}\, \mathrm{d}x^\prime =
 \hat{P}^{\mathrm{st}}(x) \left( 1\, - \, \e^{-\beta FL} \right) \e^{\beta
   A(x)}\, .
\end{align} Rearranging the terms and integrating over $0$
to $L$, Eq.~\eqref{eq:j5} leads to
\begin{align} \label{eq:j6}
\hat{J}\int_{0}^{L} \e^{-\beta A(x)} \mathrm{d}x
\int_{x}^{x+L} \frac{1}{D(x^\prime)}\, \e^{\beta A(x^\prime)}\, \mathrm{d}x^\prime = 
\left( 1\, - \,\e^{-\beta FL} \right) \int_{0}^{L} \hat{P}^{\mathrm{st}}(x) \mathrm{d}x \, ,
\end{align}
which due to the normalization yields
\begin{align}
\label{eq:j7}
\hat{J} = \frac{ \left(1-\e^{-\beta FL}\right)}
         {\int_{0}^{L}\mathrm{d}x \e^{-\beta A(x)} \int_{x}^{x+L}\mathrm{d}x^\prime
         \frac{1}{D(x^\prime)} \e^{\beta A(x^\prime)}}.
\end{align}

The general relation between {\it the stationary probability current} and {\it the
steady state particle current} ($\langle \dot{x} \rangle$) is

\begin{align}
\label{eq:current}
\langle \dot{x} \rangle = \int_{0}^{L} \hat{J}\, \mathrm{d}x
\end{align}
which implies $\langle \dot{x} \rangle = \hat{J}\, L $. Then, the
particle current reads:
\begin{align}
  \label{eq:j9}
  \langle \dot{x} \rangle = \frac{ \left(1-\e^{-\beta FL}\right)}
  { \int_{0}^{L}\frac{\mathrm{d}x}{L} \e^{-\beta A(x)} \int_{x}^{x+L}\mathrm{d}x^\prime
    \frac{1}{D(x^\prime)} \e^{\beta A(x^\prime)}}
\end{align}

Remarkably, this expression for the particle current is equivalent to
the expression obtained via the mean-first-passage-time approach
presented in Ref.~\cite{Reguera_PRL}.

\end{appendix}

\end{document}